\documentclass[preprint,authoryear,12pt]{elsarticle}

\usepackage{graphicx}

\usepackage{amssymb}

\journal{Icarus}

\begin{document}

\begin{frontmatter}



\title{Detecting the Signatures of Uranus and Neptune}

\author{Stephen R. Kane}

\address{NASA Exoplanet Science Institute, Caltech, MS 100-22, 770
  South Wilson Avenue, Pasadena, CA 91125, USA}


\begin{abstract}

With more than 15 years since the the first radial velocity discovery
of a planet orbiting a Sun-like star, the time baseline for radial
velocity surveys is now extending out beyond the orbit of Jupiter
analogs. The sensitivity to exoplanet orbital periods beyond that of
Saturn orbital radii however is still beyond our reach such that very
few clues regarding the prevalence of ice giants orbiting solar
analogs are available to us. Here we simulate the radial velocity,
transit, and photometric phase amplitude signatures of the solar
system giant planets, in particular Uranus and Neptune, and assess
their detectability. We scale these results for application to
monitoring low-mass stars and compare the relative detection prospects
with other potential methods, such as astrometry and imaging. These
results quantitatively show how many of the existing techniques are
suitable for the detection of ice giants beyond the snow line for
late-type stars and the challenges that lie ahead for the detection
true Uranus/Neptune analogues around solar-type stars.

\end{abstract}


\end{frontmatter}


\section{Introduction}
\label{introduction}

Ice giant planets, such as Uranus and Neptune, have been known to
exist for several hundred years. On July 10, 2011 one complete
Neptunian orbit will have occurred since it was discovered on
September 23, 1846 from the calculations of Le Verrier and Adams. Over
the past 15 years, the number of planets known outside of our solar
system has grown to be more than 500. Yet our time baseline combined
with the sensitivity of our measurements have not yet allowed us to
probe into regions of parameter space where we might discover ice
giants around solar analogs. Microlensing surveys are producing
insights into the frequency of giant planets beyond the ``snow line''
\citep{gou10}, but statistics regarding the orbital properties of ice
giants remains unknown and formation scenarios are deprived of the
kinds of data that now exist for Jupiter analogs in exoplanetary
systems.

The internal structure of Uranus and Neptune are known to differ quite
dramatically from that of Jupiter and Saturn \citep{for11}. Much work
has been performed on the formation mechanisms for the giant planets,
but contention still remains between the competing ideas of core
accretion and disk instability scenarios (see \citet{mat07} and
references therein). The required timescales and current orbital
configuration of the solar system giant planets favour a core
accretion model, but the formation of Uranus and Neptune are still
poorly understood \citep{ben09}. Without other examples of ice giants
to draw upon, it is difficult to understand where our ice giants
formed, timescales for their migration, and their overall
contributions to the orbital stability of the present solar system
orbits.

As we approach the heliocentric birthday of Neptune's discovery, it is
appropriate to ask how close we are to discovering a twin to Uranus or
Neptune elsewhere. With instrumentation improvements of radial
velocity surveys and time baselines now stretching out beyond the
orbit of Jupiter, we are slowly entering a phase where the detection
of Uranus and Neptune analogs is becoming more likely. Here we provide
detailed detection simulations of our giant planets from an external
perspective, concentrating on the detectability of Uranus and
Neptune. We describe the expected radial velocity and photometric
signatures and provide scaling laws for application to detection
thresholds for ice giants orbiting late-type stars which serve as far
more feasible targets in the short-term. We compare these to imaging
and astrometric experiments and assess the relative detection
prospects.


\section{Formation of Ice Giants}
\label{formation}

The formation of Uranus and Neptune in situ is problematic using core
accretion models due to the lack of disk material. A proposed solution
builds the ice giants in the same vicinity as Jupiter whereupon they
are scattered outwards once Jupiter develops its gas envelope
\citep{tho99,tho02}. Subsequent interactions with the Kuiper Belt is
likely to have both stabilized and circularized the orbits of Uranus
and Neptune to their present orbital configuration
\citep{for07,lev08}. More detailed models of the formation mechanisms
have since had success in explaining the relative chemical abundances now
seen in Uranus and Neptune \citep{dod10}.

Even though theories regarding the formation of Uranus and Neptune are
becoming more sophisticated, these still remain the only ice giants
for which we have detailed orbital information from which such models
may be constructed. A key factor is the location of the ``snow line'',
which is the radial distance from the center of a protostellar disk
beyond which water molecules can efficiently condense to form ice. For
our own system, this is approximately at the radius of the asteroid
belt, $a \sim 2.7$~AU. The formation mechanisms required for the
production of ice giants are therefore likely to be highly dependent
upon the disk mass and the production of Jupiter analogs which may
either disrupt their formation or scatter them to a larger semi-major
axis. In exoplanetary systems, we can expect to find ice giants in a
range of orbital configurations, including high eccentricities,
depending upon their formation circumstances.

\begin{figure*}
  \begin{center}
    \begin{tabular}{cc}
      \includegraphics[angle=270,width=6.5cm]{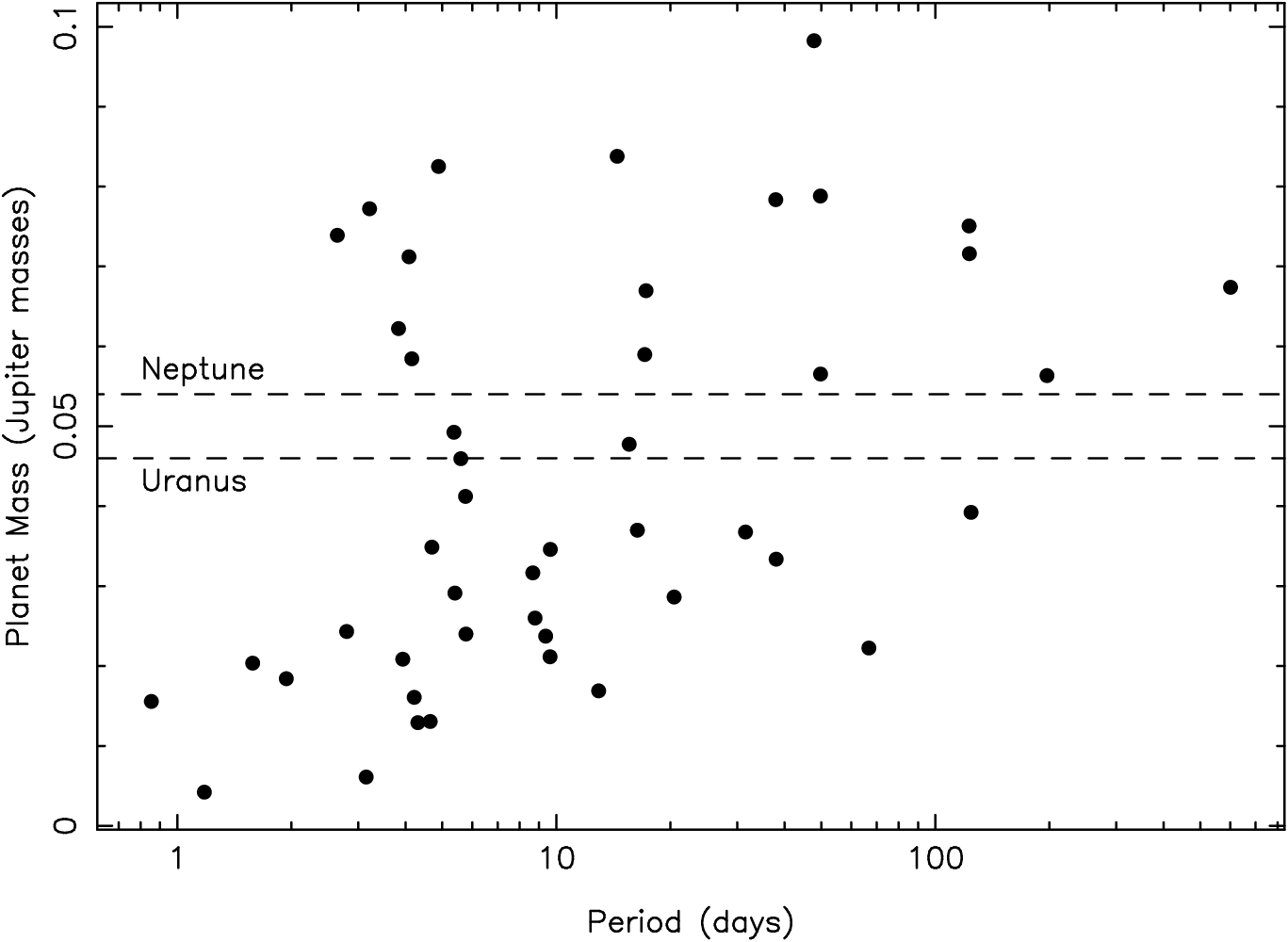} &
      \includegraphics[angle=270,width=6.5cm]{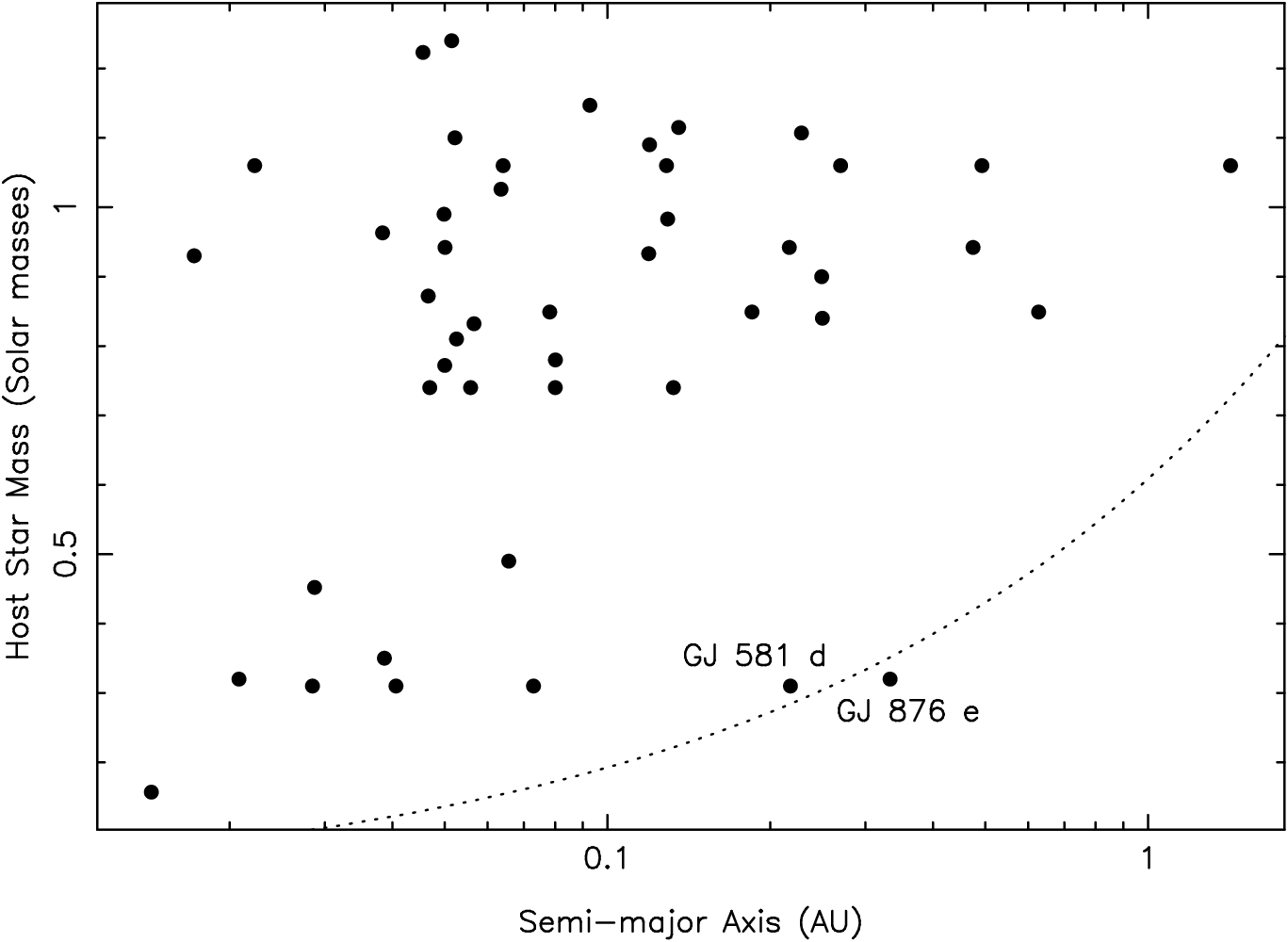}
    \end{tabular}
  \end{center}
  \caption{Known radial velocity planets with masses less than 0.1
    Jupiter masses. Left: planet masses as a function of orbital
    period with Neptune and Uranus indicated by dashed lines. Right:
    Host star mass as a function of semi-major axis where the
    predicted snow-line is shown as a dotted line.}
  \label{masses}
\end{figure*}

\citet{ida05} approximate the location of the snow line as a function
of stellar mass, $M_\star$, by $a_{\mathrm{ice}} = 2.7 (M_\star /
M_\odot)^2$. \citet{ken06} further develop the time-dependence of the
snow line location in the context of low-mass stars. This work is
extended by \citet{ken08}, in which the location of the snow line is
generalized to stars of various masses. Using these models, one can
investigate whether true ice giants, both in terms of formation and
composition, have already been detected around low-mass stars. Shown
in Figure \ref{masses} are the known radial velocity exoplanets with
masses in the vicinity of Uranus and Neptune. These data were
extracted using the Exoplanet Data Explorer\footnote{\tt
  http://exoplanets.org/} and are current as of 1st January 2011. The
left panel shows the minimum planet masses ($M_p \sin i$, where $i$ is
the orbital inclination) as a function of the orbital period, where
the dashed lines indicate the masses of Uranus and Neptune for
comparison. There have been a number of planets discovered with masses
lower than the ice giants of our solar system, but these are
preferentially found closer to their parent stars since this produces
a larger radial velocity signature. The right panel shows the masses
of the host stars for these planets as a function of the semi-major
axis of the planetary orbits. The dotted line indicates the location
of the snow line using the approximation of \citet{ida05}. The planet
located just inside the snow line is GJ~581~d, a planet substantially
less massive than Neptune ($\sim 7$ Earth masses) and whose eccentric
orbit takes it either side of the snow line as it passes between
apastron and periastron \citep{may09}. A more analogous ice giant
example is the case of GJ~876~e, shown just outside of the snow
line. This planet has a mass between that of Uranus and Neptune and an
almost circular orbit \citep{riv10}. The existence of several more
massive gas giants within the system at smaller orbital radii may
indicate that migration during and after the formation of the ice
giant also took place within this system \citep{bos06,bos10}.


\section{Radial Velocity Signatures}


\subsection{The Signatures of Uranus and Neptune}
\label{rvsig}

The simulated data assume a precision of 10~cm/s with a cadence of one
measurement per year over a complete Neptune orbit. We assume a
cadence of one measurement per year over a complete Neptune orbit
(164.8 years). The long-term stability of precision radial velocity
instruments presents a significant challenge to achieving this. Recent
examples include 10 years of observations of HD~185144 that were
carried out using Keck/HIRES by \citet{wri08} from which they achieved
an RMS scatter of the radial velocity variations of less than 2.5~m/s
over the entire period. The HARPS instrument has already demonstrated
stability of 1--2~m/s over the timescale of months with further
improvements expected in the near future \citep{pep08}. A long-term
solution is the use of laser frequency combs \citep{li08,mur07} which
can achieve long-term stability by stabilizing the frequency modes
with the absolute frequencies of atomic clocks \citep{ste08,ude02}.

Studies of intrinsic stellar variability and solar data show that many
inactive stars exist, at least to the precision at which they have
been studied \citep{wri05}, and that many types of activity may be
circumvented \citep{boi11,lag10,meu10}.  For example, star spot
activity produces both radial velocity and photometric jitter. Star
spots tend to persist for no longer than a few weeks for main sequence
stars \citep{str09} but their frequency rises and falls over longer
timescales. The Sun has cycles of duration 11, 22, and 87 years for
the Schwabe, Hale, and Gleissberg cycles respectively. These cycles
can be identified and removed through simultaneous photometric and
radial velocity monitoring and also through activity indicators for
the star \citep{mak09}. The star spot activity can also be tied to the
stellar rotation periods which has been performed for numerous known
exoplanet host stars \citep{sim10}.

\begin{figure}
  \begin{center}
    \begin{tabular}{cc}
      \includegraphics[angle=270,totalheight=3.3cm]{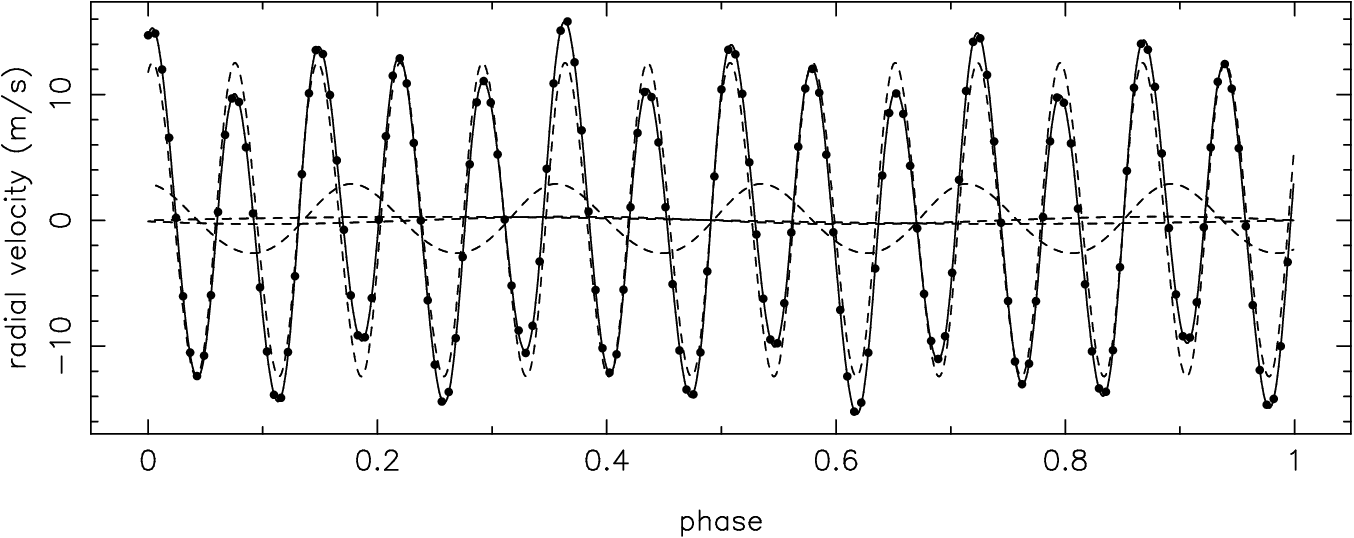} &
      \includegraphics[angle=270,totalheight=3.3cm]{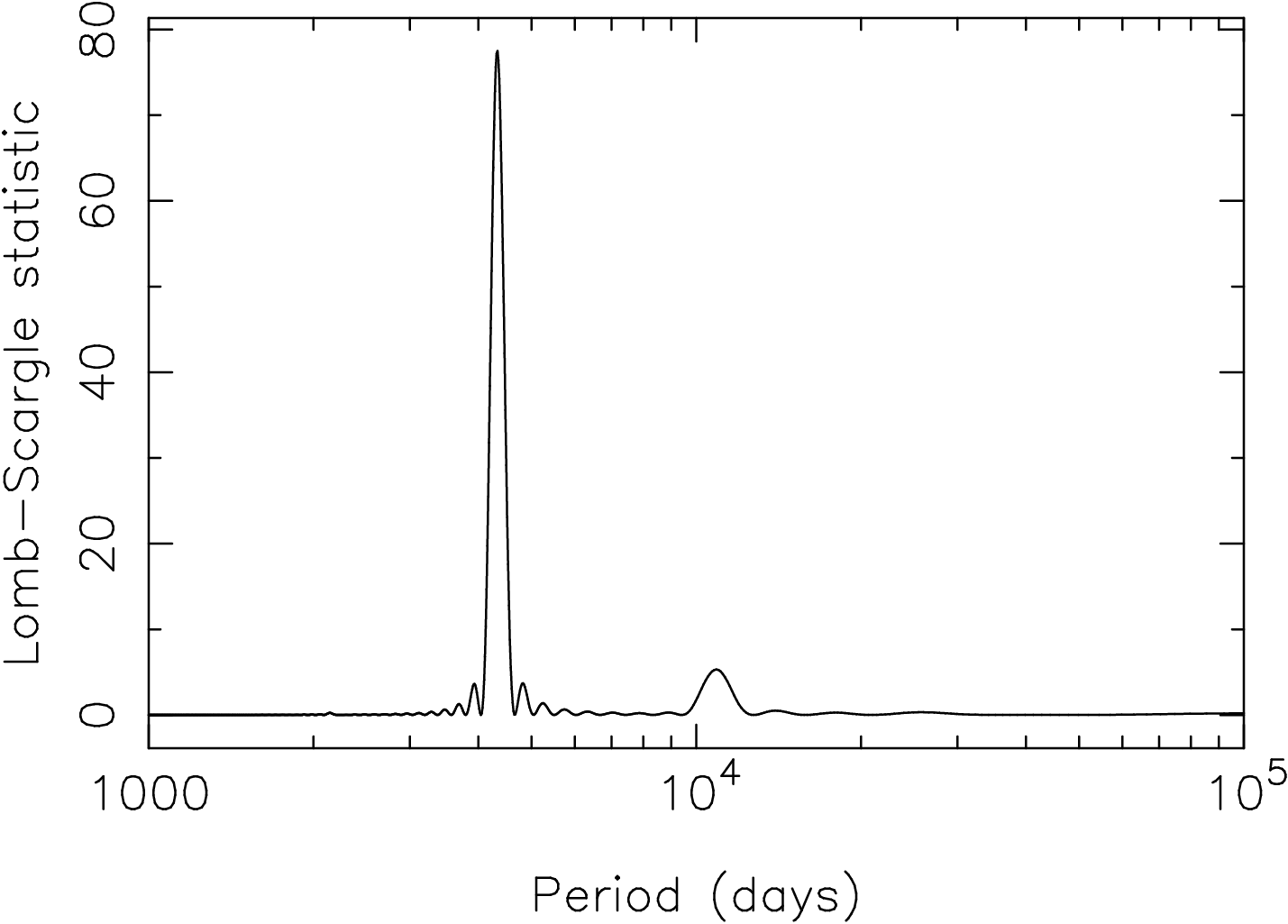} \\
      \includegraphics[angle=270,totalheight=3.3cm]{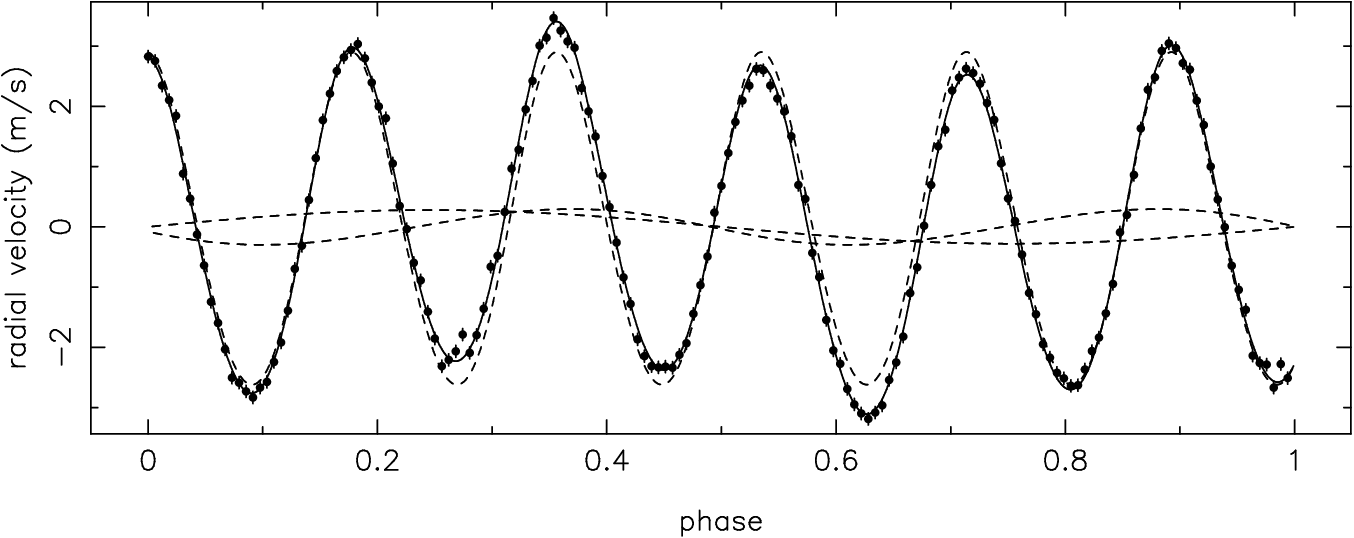} &
      \includegraphics[angle=270,totalheight=3.3cm]{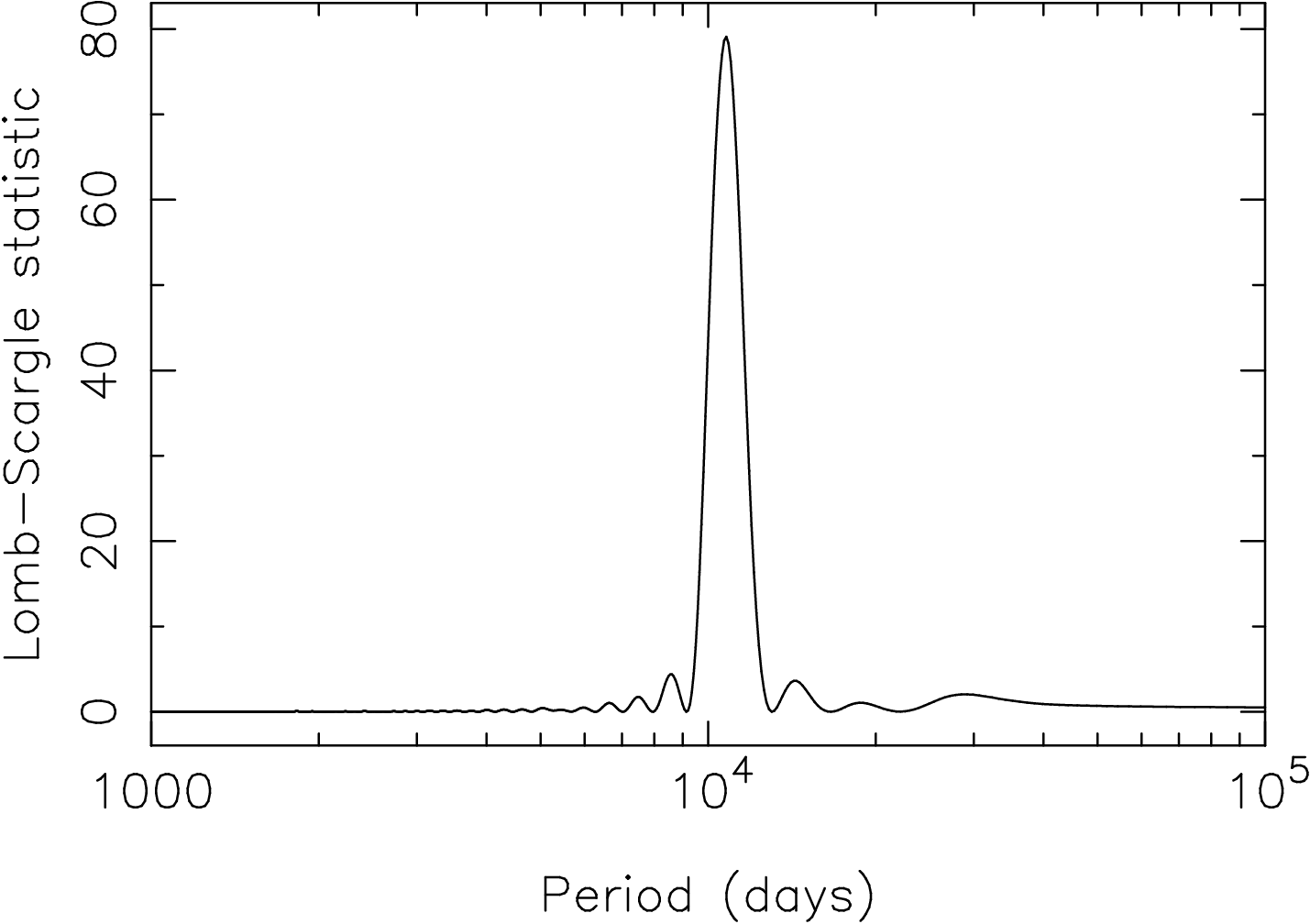} \\
      \includegraphics[angle=270,totalheight=3.3cm]{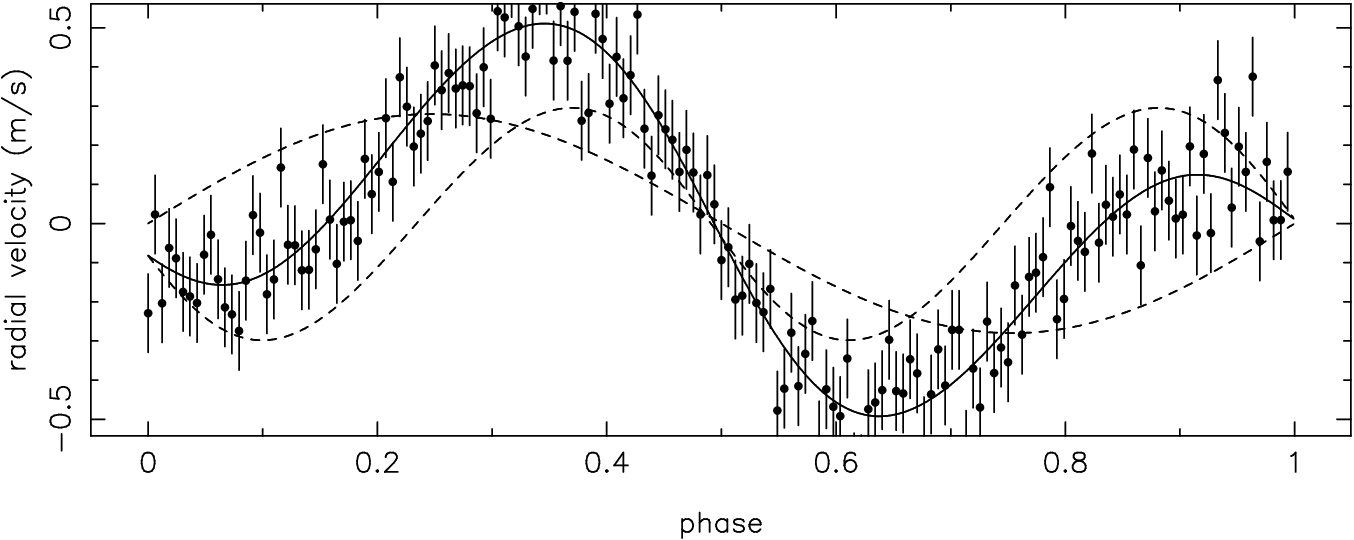} &
      \includegraphics[angle=270,totalheight=3.3cm]{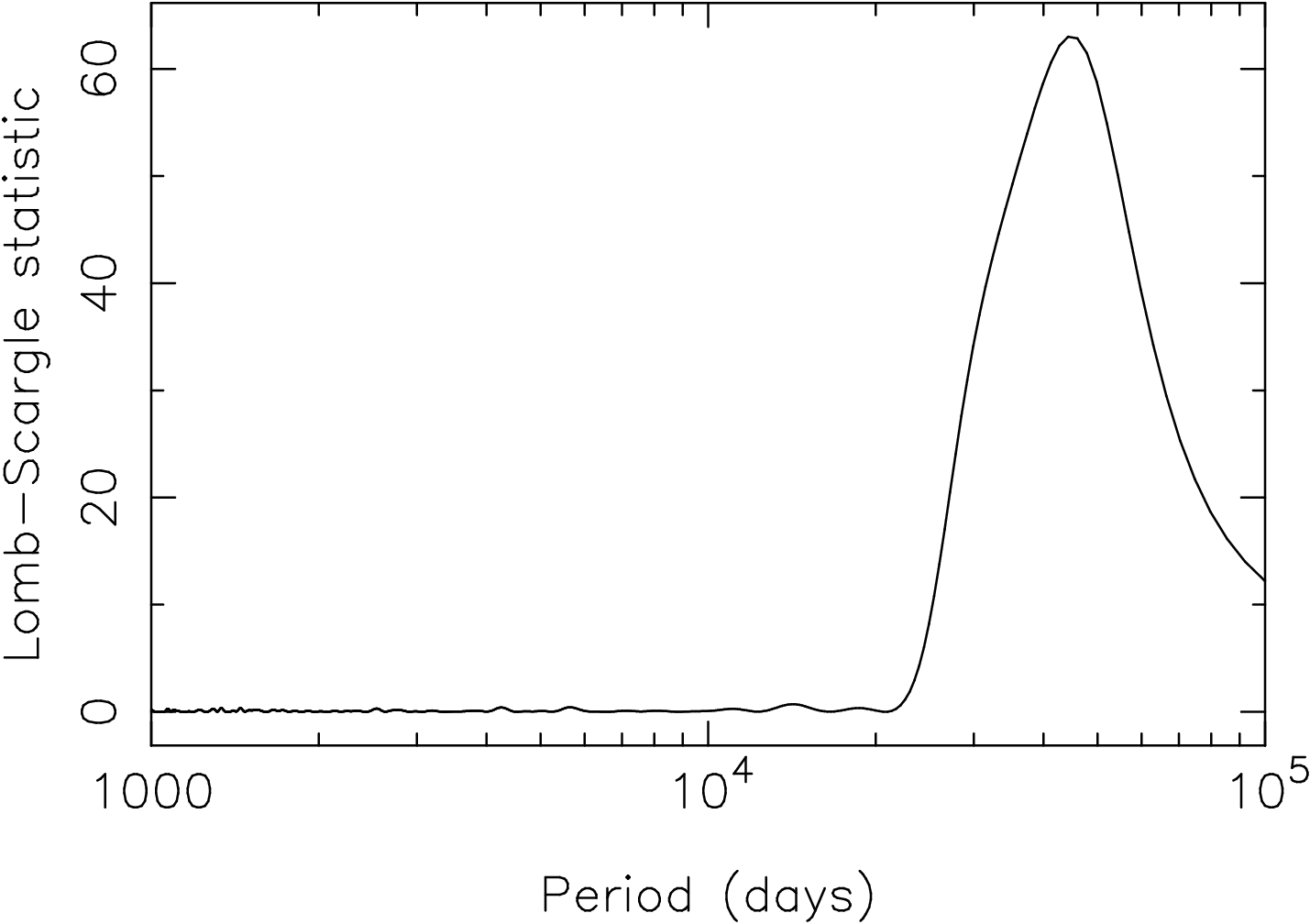}
    \end{tabular}
  \end{center}
  \caption{Combined radial velocity signature (solid lines) and
    individual planetary signatures (dashed lines) for the giant
    planets of our solar system, along with simulated radial velocity
    measurements with a cadence of $\sim 1$~year. The periodograms in
    the right panels show the Fourier analysis of the combined
    signals. The top panel includes Jupiter, Saturn, Uranus, and
    Neptune. The middle panel includes Saturn, Uranus, and
    Neptune. The bottom panel includes Uranus and Neptune. In each
    case, the orbital phase is normalized to the orbital period of
    Neptune.}
  \label{rvss1}
\end{figure}

Figure \ref{rvss1} shows the result of the simulation for the solar
system giant planets, first with all four planets and then
systematically removing the signature of each planet in order of
increasing orbital period until the combined signal of Uranus and
Neptune remain. Data for the solar system planets were extracted from
the JPL HORIZONS System\footnote{\tt
  http://ssd.jpl.nasa.gov/?horizons}. The semi-amplitude of the radial
velocity signal, $K$, is given by
\begin{equation}
  K = \left( \frac{2 \pi G}{P} \right)^{1/3} \frac{M_p \sin
    i}{(M_p + M_\star)^{2/3}} \frac{1}{\sqrt{1-e^2}}
  \label{rveqn}
\end{equation}
where $P$ is the period, $i$ is the inclination of the planetary
orbit, $e$ is the orbital eccentricity, and $M_p$ is the planetary
mass. The inclination of the orbital plane to the observer
line-of-sight is assumed to be edge-on ($i = 90^\circ$). The
semi-amplitude for each of the planets under these conditions is shown
in Table \ref{rvtab}.

\begin{table}
  \begin{center}
    \caption{Radial velocity parameters of the giant planets.}
    \label{rvtab}
    \begin{tabular}{@{}lcccc}
      \hline
      Planet  & $P$ (days) & $M_p$ ($M_J$) & $e$   & $K$ (m/s) \\
      \hline
      Jupiter &  4332.820  & 1.000         & 0.049 & 12.47 \\
      Saturn  & 10755.698  & 0.299         & 0.056 &  2.76 \\
      Uranus  & 30687.153  & 0.046         & 0.044 &  0.30 \\
      Neptune & 60190.029  & 0.054         & 0.011 &  0.28 \\
      \hline
    \end{tabular}
  \end{center}
\end{table}

In each panel of Figure \ref{rvss1} the dashed lines indicate the
individual planetary signals and the solid line represents the
combined signal. Next to each of the panels is a Lomb-Scargle (L-S)
periodogram \citep{lom76,sca82} from a Fourier analysis of the
data. The Jupiter and Saturn signatures are clearly detected in their
associated periodograms. The semi-amplitude of the Uranus and Neptune
signatures are almost identical due to the lower mass of Uranus being
compensated by its smaller orbital period. The Fourier disentanglement
of the ice giant signatures is ambiguous with these data, although
Keplerian orbital fitting will be more successful at extracting the
independent parameters. Since the number of cycles is very limited, an
approach which utilizes the Maximum Entropy Method (MEM) may be more
helpful since this is relatively efficient in detecting frequency
lines with few assumptions regarding the initial estimates of the fit
parameters.

An important factor in the detection process is the eccentricity of
the orbits. As shown in Table \ref{rvtab}, the eccentricities for the
solar system giant planets are all relatively low, less than
0.06. Many of the known exoplanets have eccentric orbits and the
scattering of ice giants to larger semi-major axes, as described in
Section \ref{formation}, may contribute to this if other stabilizing
influences are not present. Biases against the detection of non-zero
eccentricities have been previously noted by \citet{kan07} and
\citet{oto09} which may be resolved through the use of Keplerian
periodograms and vigorously determined false-alarm thresholds
\citep{cum04}.


\subsection{Detection vs Time}

\begin{figure}
  \begin{center}
    \begin{tabular}{cc}
      \includegraphics[angle=270,totalheight=3.3cm]{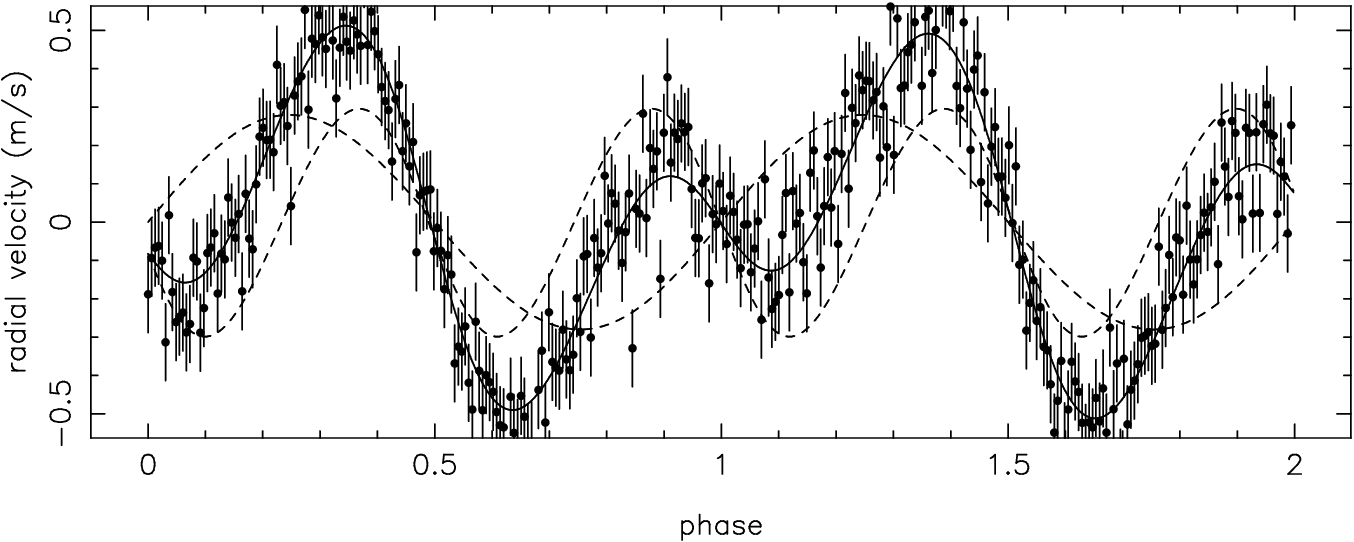} &
      \includegraphics[angle=270,totalheight=3.3cm]{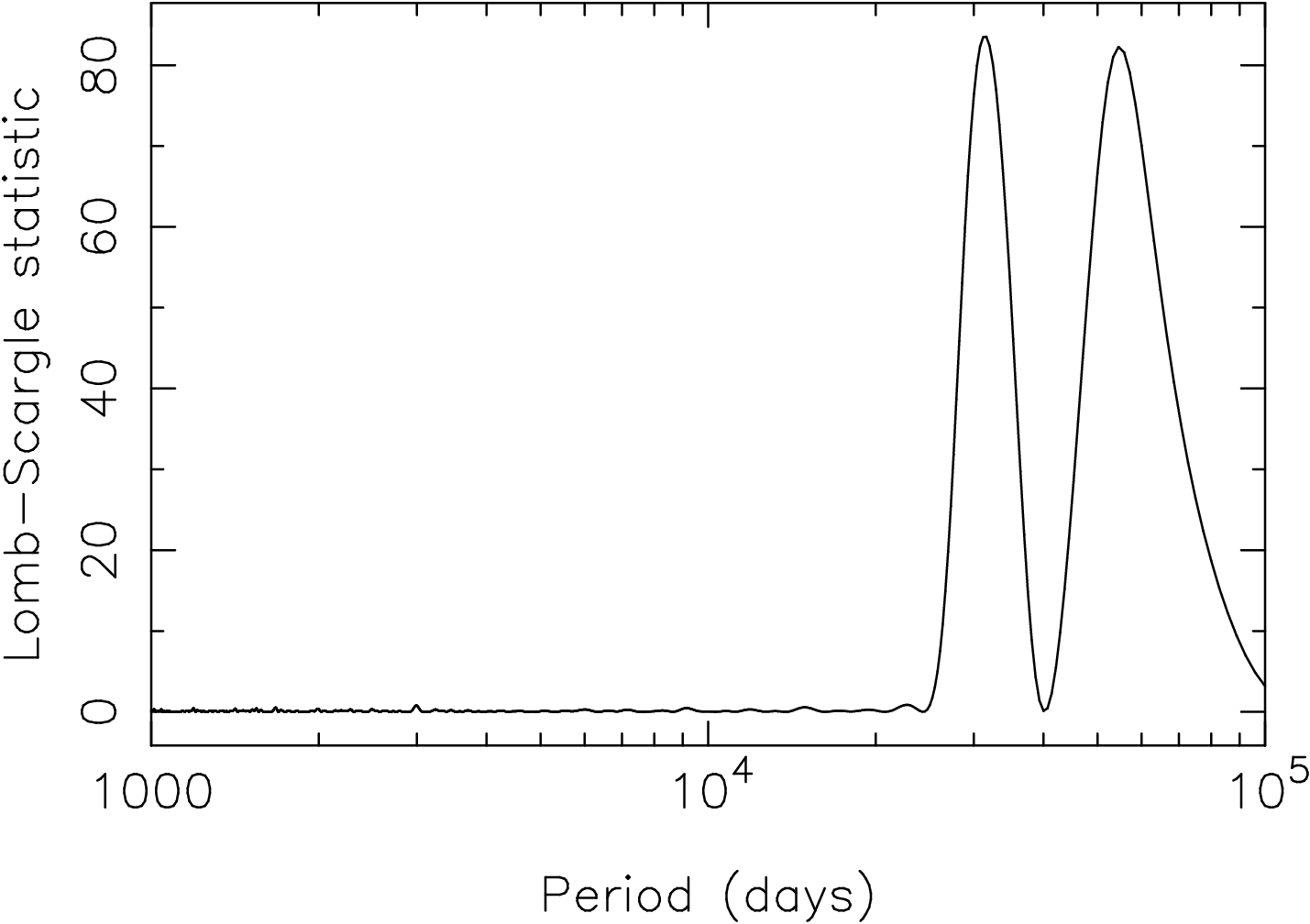} \\
    \end{tabular}
  \end{center}
  \caption{Combined radial velocity signature (solid lines) and
    individual planetary signatures (dashed lines) for Uranus and
    Neptune, along with simulated radial velocity measurements, after
    monitoring the host star for two Neptunian orbital periods. The
    periodogram in the right panel shows the Fourier analysis of the
    combined signal.}
  \label{rvss2}
\end{figure}

The ambiguous detection of Uranus and Neptune described in the
previous section is partially due to the similar amplitudes of the
radial velocity signals, but mostly due to the inadequate orbital
periods monitored to produce a strong power in the Fourier
spectrum. The dependence of a detection on the number of radial
velocity measurements has been previously studied by
\citet{kan07}. Figure \ref{rvss2} shows how this ambiguity resolves
when the system is monitored for two complete Neptunian orbital
periods, still with a cadence of $\sim 1$~year. The resolution of the
two distinct planetary signatures in the Fourier spectrum allows a
Keplerian fit to easily recover the orbital parameters of the two
planets.

\subsection{Scaling to Low-mass Stars}
\label{scaling}

A far more accessible parameter regime to probe in the near future is
that surrounding the low mass stars. By combining the approximate location
of the snow line \citep{ida05} with Equation \ref{rveqn}, we
produce an expression for the expected radial velocity semi-amplitude
at the location of the snow line as a function of mass. This is given
by
\begin{equation}
  K = 18129 \times \frac{M_p \sin i}{M_\star^{5/6} (M_p +
    M_\star)^{2/3}} \frac{1}{\sqrt{1-e^2}}
  \label{rveqn_lms}
\end{equation}
where the mass of the planet and star are in units of solar masses for
convenience. This expression may be considered an upper limit for the
radial velocity signatures of possible ice giant analogues.

For example, if we placed a Neptune-mass planet at the snow line of a
solar-mass star, the period of the orbit will be $\sim 1620$~days and
the semi-amplitude of the radial velocity signature will be $\sim
0.93$~m/s. However, if the Neptune-mass planet were orbiting at the
snow line of a late M dwarf (0.3~$M_\odot$) the period becomes $\sim
80$~days and the semi-amplitude of the radial velocity signature rises
to $\sim 5.7$~m/s. This detection scenario is not only achievable, but
has been significantly surpassed by current radial velocity surveys,
indicating that we are now starting to build a picture of the
frequency of Uranus/Neptune analogues around M dwarfs if not higher
mass stars.


\subsection{Linear Trends}

According to the Exoplanet Data Explorer, 37 of the known radial
velocity systems have measured linear trends in the residuals of the
Keplerian orbital fitting. Of these, 9 are known to have a stellar
companion which may in some cases be the sources of the trend. The
linear trends have amplitudes which lie in the range
0.0014--0.14~m/s/day. This is equivalent to 0.511--51.1~m/s/year which
is already enough to exclude the detection of Uranus/Neptune analogues
depending upon the mass of the host star and the timescale over which
the trend has already been observed. Consider the case of HD~44219b
\citep{nae10}, which has a minimum mass of $0.58 M_J$ and is in a
472~day orbit around a G2V star. The Keplerian solution for this
planet includes a quadratric drift of −0.76390~m/s/year$^2$ which
places this signature at the threshold of the expected semi-amplitude
for a Neptune located at the snow line for a solar-type star, as
described in Section \ref{scaling}. These kinds of trends will
preferentially be caused by large mass objects since these produce
larger signatures, but a greater time baseline is needed to resolve
these potential ice giant detections.


\section{Transit Signatures}

The discovery of exoplanets using the transit technique has become
increasingly dominant amongst the various detection methods. Examples
of major contributors to the ground-based discovery of transiting
exoplanets are the Hungarian Automated Telescope Network (HATNet)
\citep{bak04} and SuperWASP \citep{pol06}. Ground-based surveys tend
to be highly restricted in the period-space which they are able to
sample, mostly due to the observational window function
\citep{von09}. This is siginificantly improved by conducting such
surveys from space, where the current major contributors are the
Kepler mission \citep{bor10} and the CoRoT mission \citep{bar08}.
The recent results released by the Kepler mission by \citet{bor11}
demonstrate that this mission is more than capable of detecting the
signatures of Neptune-radii planets, though it will take further more
time to detect these beyond the snow line for those stars in the
Kepler Input Catalog \citep{bro11}.

The change in flux from the host star due to the transit of a planet is
\begin{equation}
  \frac{\Delta F}{F} = \left( \frac{R_p}{R_\star} \right)^2
\end{equation}
where $R_p$ and $R_\star$ are the radii of the planet and star
respectively. The probability that a transit will be observable is
given by
\begin{equation}
  P_t = \frac{(R_p + R_\star)(1 + e \cos (\pi/2 - \omega))}{a (1 -
    e^2)}
\end{equation}
where $\omega$ and $a$ are the argument of periastron and semi-major
axis of the orbit respectively \citep{kan08}. Applying these to
Neptune yield a transit depth of 0.13\% and a transit probability of
only 0.015\%. Even though the photometric precision required is
already achieved, it is difficult to imagine a survey that would
monitor a star for 184 years with such small prospects for detecting a
transit! If we consider a Neptune-radius planet at the snow line of a
solar-type star, then the depth remains the same but the probability
increases to 0.17\%. As noted in Section \ref{scaling}, the orbital
period in this case will be $\sim 1620$~days. Thus the Kepler mission
may well detect single transits from such ice giants during the
mission duration. Scaling to low-mass stars, a Neptune-radius planet
located at the snow line of a late M dwarf (0.3~$R_\odot$) results in
a depth of 1.41\% and a probability of 0.57\%. These values place such
planets well within the reach of Kepler and could even be detected by
targetted transit searches of low-mass stars, such as the MEarth
Project \citep{cha09}.


\section{Photometric Phase Signatures}

As an exoplanet orbits the host star, the changing phase exhibits a
distinct photometric signature. The photometric phase signatures of
giant planets have been described in detail by \citet{sud05} and
\citet{cah10}. However, the flux ratio of the planet to the host star
is small compared to typical current photometric precisions and has
presented a major hinderance to the realization of detecting such
signatures. There is also the issue of stellar variability to contend
with (see Section \ref{rvsig}).

Investigating the giant planets of our solar system presents a
significant challenge for even current space-based
capabilities. \citet{kan10} and \citet{kan11} calculate the predicted
amplitudes for giant planets in long-period orbits but only find
detectable signatures during periastron passage for those in eccentric
orbits. The flux ratio of the planet to the host star is defined as
\begin{equation}
  \epsilon(\alpha,\lambda) \equiv
  \frac{f_p(\alpha,\lambda)}{f_\star(\lambda)}
  = A_g(\lambda) g(\alpha,\lambda) \frac{R_p^2}{r^2}
  \label{fluxratio}
\end{equation}
where $\alpha$ is the phase angle (defined to be zero at superior
conjunction), $\lambda$ is the wavelength, $A_g$ is the geometric
albedo, $g$ is the phase function, and $r$ is the star--planet
separation. For the phase function, we adopt the ``Hilton'' function
utilised by \citet{kan10} which is based upon Pioneer observations of
Venus and Jupiter. Table \ref{phatab} contains the relevant orbital
and physical parameters for this simulation, along with the predicted
maximum flux ratios for each of the planets. The geometric albedos of
the planets are well known and were also extracted from the JPL
HORIZONS System.

\begin{table}
  \begin{center}
    \caption{Phase variation parameters of the giant planets.}
    \label{phatab}
    \begin{tabular}{@{}lccccc}
      \hline
      Planet  & $P$ (days) & $R_p$ ($R_J$) & $e$   & $A_g$ & $\epsilon_{\mathrm{max}} (10^{-9})$ \\
      \hline
      Jupiter &  4332.820  & 1.0000        & 0.049 &  0.52 & 4.845 \\
      Saturn  & 10755.698  & 0.8430        & 0.056 &  0.47 & 0.883 \\
      Uranus  & 30687.153  & 0.3575        & 0.044 &  0.51 & 0.037 \\
      Neptune & 60190.029  & 0.3464        & 0.011 &  0.41 & 0.013 \\
      \hline
    \end{tabular}
  \end{center}
\end{table}

\begin{figure}
  \includegraphics[angle=270,totalheight=3.8cm]{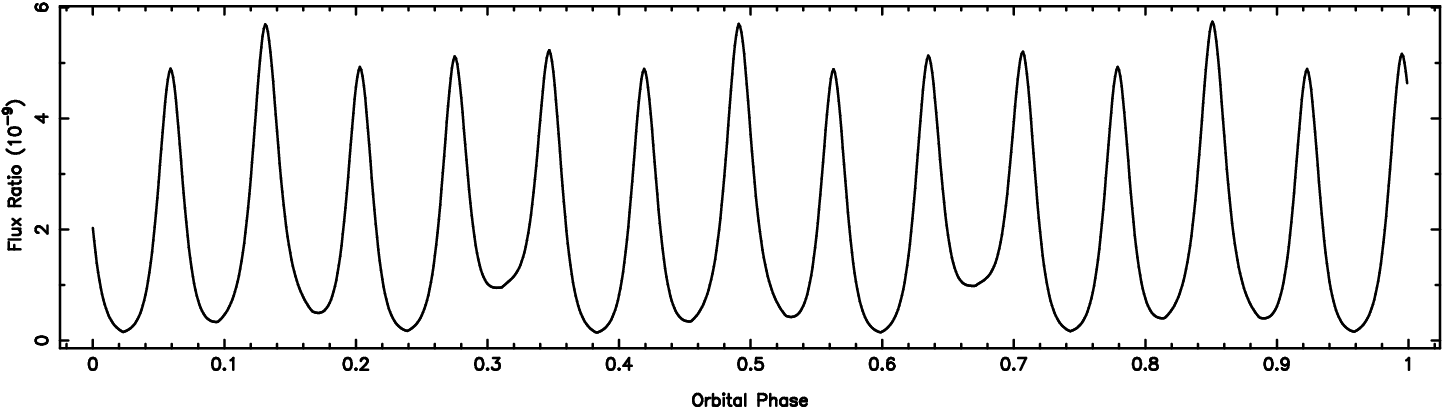}
  \includegraphics[angle=270,totalheight=3.8cm]{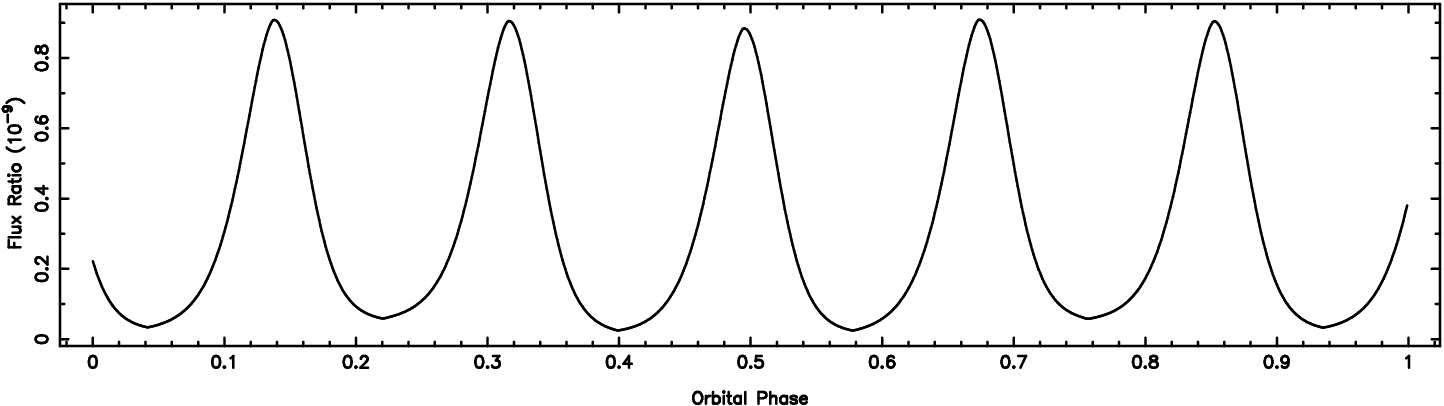}
  \includegraphics[angle=270,totalheight=3.8cm]{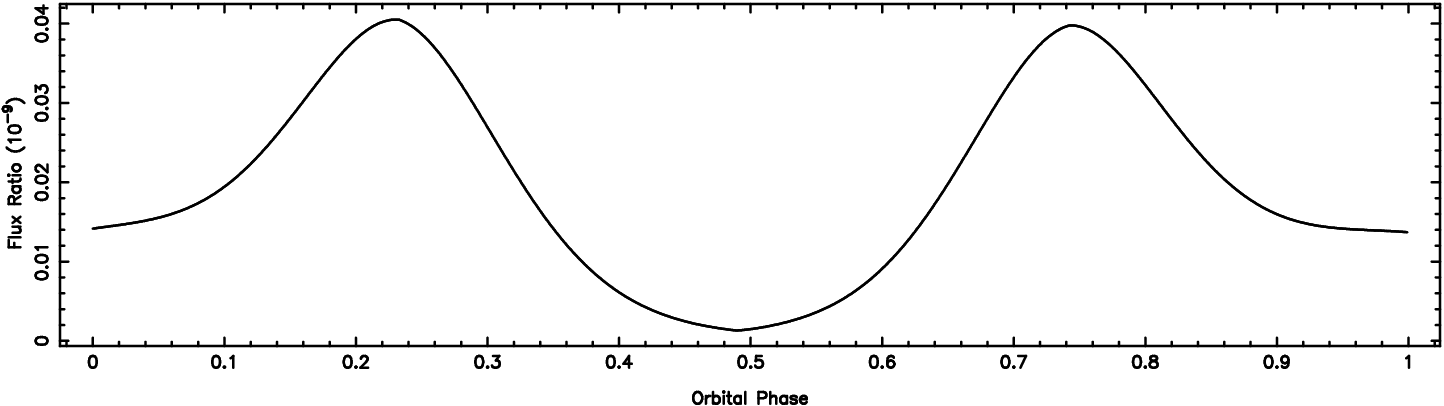}
  \includegraphics[angle=270,totalheight=3.8cm]{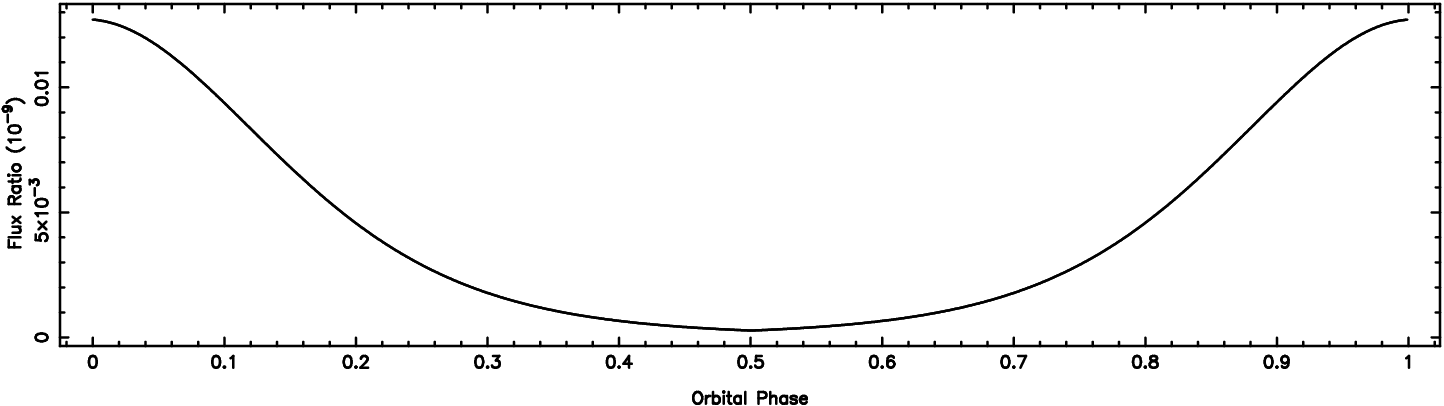}
  \caption{Combined phase signature for the giant planets of our solar
    system. The top panel includes Jupiter, Saturn, Uranus, and
    Neptune. Each successive panel removes the signature of a planet
    in order of their semi-major axis, ending with the phase signature
    of Neptune in the bottom panel. In each case, the orbital phase is
    normalized to the orbital period of Neptune.}
  \label{phass}
\end{figure}

Figure \ref{phass} shows the predicted phase signature for the solar
system giant planets at optical wavelengths. The top panel includes
all four planets. With each successive panel, the planetary signatures
are removed in order of increasing semi-major axis until only the
signature of Neptune remains. There are several key aspects to note
here. Firstly, the amplitude of these signatures, shown in the figure
and also in Table \ref{phatab}, are below what is currently
accessible. For comparison, the ellipsoidal variations detected by
\citet{wel10} for the planet HAT-P-7b using Kepler data was of
amplitude $3.7 \times 10^{-5}$.  Secondly, this does not take into
account ring structures such as the prominent ring system of Saturn
\citep{dyu05}, and so the phase signature is underestimated for some
planets.  Thirdly, these signatures are significantly lower than the
solar variability, which was compared to the photometric variability
found amongst Kepler stars by \citep{bas10}. The question of whether
one could maintain precise photometric stability over long timescales
is a related problem which is detector dependent. Fourthly, this does
not include contamination from the terrestrial planets. As shown in
Equation \ref{fluxratio}, the inverse relation of the star--planet
separation will generally dominate over differences in planetary
radii. Venus, for example, has an albedo of 0.65 which, combined with
the much smaller semi-major axis, leads to a maximum flux ratio of
$2.06 \times 10^{-9}$ or 43\% for that of Jupiter. Thus, the possible
presence of much smaller terrestrial planets can not be ignored.

By using the same examples as those described in Section
\ref{scaling}, we can test how this seemingly dire situation were to
change if Neptune were to be found in different situations. If we
place Neptune at the snow line of a solar-type star ($P = 1620$~days),
then the predicted maximum flux ratio increases to $1.57 \times
10^{-9}$, just slightly lower than that of Venus. If however, Neptune
were to be located at the snow line of a late M dwarf ($P = 80$~days),
then the predicted maximum flux ratio becomes $1.95 \times
10^{-7}$. This is two orders of magnitude below the flux ratio
detected in HAT-P-7 by \citet{wel10}. However, such flux ratios are
not beyond the reach of planned coronographic missions and
thirty-meter class telescopes and may provide an opportunity to not
only detect an ice giant but to characterize the atmosphere.


\section{Astrometry and Imaging}

The limits on using astrometry for detecting outer planets have been
explored by \citet{eis01}, and further investigated by combining with
radial velocity data by \citet{eis02}. Direct imaging of exoplanets
has had recent success, such as the discovery of the planetary system
orbiting the nearby star HR~8799 with semi-major axes as small as
14.5~AU \citep{mar08,mar10}. The techniques of astrometry and imaging
are distinct from the previous two methods investigated as they (a)
are more sensitive to long-period orbits and (b) depend upon the
distance to the system. They are further distinctive in that they are
more sensitive to face-on orbits rather than edge-on orbits for which
the radial velocity, transit, and (often) phase variation techniques
typically favour. An addition factor for the imaging technique is the
dependence upon the age of the system since young planets ($< 1$~Gyr)
will have self-luminous properties.

The amplitude of the astrometric signature is given by
\begin{equation}
  \Delta \theta = \left( \frac{M_p}{M_\star} \right) \left(
  \frac{a}{d} \right)
\end{equation}
According to the Exoplanet Data Explorer, the distribution of
distances to the host stars for the known radial velocity planets
peaks at $\sim 35$~pcs, due largely to the observational biases in the
target selection. Neptune viewed from this distance has an angular
separation from the Sun of 0.86~arcsec but a predicted astrometric
signature of 44~$\mu$arcsec. However, a Neptune-mass planet located at
the snow line of a late M dwarf has an angular separation of
0.007~arcsec and a predicted astrometric signature of
1.2~$\mu$arcsec. Thus, if such techniques are to be used for
attempting the detection of ice giants, actual solar analogues are
preferred targets over late-type stars. Current and future
coronographic experiments may play a role in probing the regime of ice
giants as sensitvities and angular resolutions
increase. \citet{bei10}, for example, tabulates the inner working
angle of selected future ground-based imaging instruments which lie in
the range 0.03--0.17~arcsec, compared to 0.035--0.850~arcsec for James
Webb Space Telescope (JWST) instruments.


\section{Conclusions}

The formation and subsequent migration of ice giants appear to have
played a major role in the overall formation and evolution of the
outer solar system. Whether this is generally true of other solar-type
systems and the frequency of such ice giants remains to be seen. The
current sample of exoplanets is probing deeper into this regime in
terms of mass but needs a longer time baseline to establish the period
sensitivity to detect true Uranus/Neptune analogues. As existing
radial velocity surveys continue to monitor known exoplanet hosting
stars and simultaneously improve instrument precisions, the growing
sample will slowly push further into the regime of the ice giants. The
regime is also being explored by the Kepler and CoRoT missions using
the transit technique from space where it is the mission duration and
not the observational window function that constrains the detection of
long-period planets. Detecting photometric phase variations poses
significant technological challenges in the short-term and
astrophysical challenges including intrinsic stellar variability and
contamination from interior terrestrial planets.

The current situation favours exploring Neptune-mass planets beyond
the snow line of late-type stars, where the required precision and
timescales are significantly less, and will reveal the formation
mechanisms in these environments. Results from the Kepler mission are
allowing the study of transit timing variations which are sensitive
enough to detect longer-period planets in those systems. In addition,
continuing microlensing surveys have enough current sensitivity to
provide a statistical census of ice giants through which we can
determine the frequency of these objects, though not necessarily their
orbital and physical characteristics.  It has been demonstrated by
\citet{col10} that precision photometry of $< 0.05$\% can be achieved
with large telescopes through the use of narrow-band filters.
Simulations of current and planned coronagraphs by \citet{bei10}
indicate that long-term stability may indeed be possible for exoplanet
searches. Future large ground-based telescopes, such as the European
Extremely Large Telescope (E-ELT), the Thirty Meter Telescope (TMT),
and the Giant Magellan Telescope (GMT), as well as space-based
coronographic missions, may thus provide the needed precision for more
direct detections via phase variation signatures and imaging.


\section*{Acknowledgements}

The author would like to thank Kaspar von Braun, Jason Wright, Suvrath
Mahadevan, and Chas Beichman for several useful discussions. I would
also like to thank the referees, whose comments helped to improve the
quality of this work. This research has made use of the Exoplanet
Orbit Database and the Exoplanet Data Explorer at exoplanets.org.


\end{document}